\documentclass[acmsmall]{acmart}

\usepackage{multirow}
\usepackage{graphicx}
\usepackage{subfigure}
\usepackage{booktabs}
\usepackage{array}
\AtBeginDocument{%
  }

\setcopyright{acmlicensed}
\copyrightyear{2024}
\acmYear{2024}
\acmDOI{XXXXXXX.XXXXXXX}

\begin{document}

\title{Can LLM Assist in the Evaluation of the Quality of Machine Learning Explanations?}

\author{Bo Wang}
\email{bo.wang-11@student.uts.edu.au}
\affiliation{%
 \institution{Data Science Institute, University of Technology Sydney}
 \city{Sydney}
 \country{Australia}
}

\author{Yiqiao Li}
\email{yiqiao.li@uts.edu.au}
\affiliation{%
 \institution{Data Science Institute, University of Technology Sydney}
 \city{Sydney}
 \country{Australia}
}

\author{Jianlong Zhou}
\email{jianlong.zhou@uts.edu.au}
\affiliation{%
 \institution{Data Science Institute, University of Technology Sydney}
 \city{Sydney}
 \country{Australia}
}

\author{Fang Chen}
\email{fang.chen@uts.edu.au}
\affiliation{%
 \institution{Data Science Institute, University of Technology Sydney}
 \city{Sydney}
 \country{Australia}
}






\begin{abstract}
  EXplainable machine learning (XML) has recently emerged to address the mystery mechanisms of machine learning (ML) systems by interpreting their 'black box' results. Despite the development of various explanation methods, determining the most suitable XML method for specific ML contexts remains unclear, highlighting the need for effective evaluation of explanations. The evaluating capabilities of the Transformer-based large language model (LLM) present an opportunity to adopt LLM-as-a-Judge for assessing explanations. In this paper, we propose a workflow that integrates both LLM-based and human judges for evaluating explanations. We examine how LLM-based judges evaluate the quality of various explanation methods and compare their evaluation capabilities to those of human judges within an iris classification scenario, employing both subjective and objective metrics. We conclude that while LLM-based judges effectively assess the quality of explanations using subjective metrics, they are not yet sufficiently developed to replace human judges in this role.
\end{abstract}


\maketitle

\section{Introduction}

Machine learning (ML) systems have seen significant growth in popularity due to their increasing problem-solving capabilities, which have been applied across diverse fields such as medical healthcare~\cite{job2024optimal}, biology analysis~\cite{molho2024deep}, and fraud detection~\cite{chen2024credit}. However, these systems often operate as 'black boxes,' lacking transparency and preventing users from understanding the rationale behind their decisions. To address this issue, researchers have developed eXplainable ML (XML) techniques, which aim to provide interpretability and insight into these opaque models, enhancing their trustworthiness and reliability.

Various methods have been proposed in the field of XML, including SHAP~\cite{lundberg2017unified}, LIME~\cite{ribeiro2016should}, and similarity-based explanations~\cite{charpiat2019input}. Despite the advancement of these methods, evaluating the quality of explanations remains a complex and unresolved issue~\cite{zhou2021evaluating, nauta_anecdotal_2023}. Thus, there is a pressing need to evaluate the quality of explanations and identify the most appropriate XML methods for practical applications. 

Existing research has explored various approaches to evaluating explainability, encompassing functionality-grounded, application-grounded, and human-grounded methods~\cite{doshi2017towards}. However, these evaluation techniques have notable limitations. Functionality-grounded evaluations rely on quantitative metrics that often fail to capture the nuances of human perception. Meanwhile, application-grounded and human-grounded evaluations, which involve experiments with experts and non-experts, are resource-intensive, requiring significant time, cost, and ethical approvals. Alternatively, researchers have investigated the use of Transformer models, specifically large language models (LLMs) as judges, positioning them as alternatives to human evaluators~\cite{hu2024unveiling, peng2024survey, raju2024constructing, shankar2024validates}. Therefore, employing the Transformer-based LLMs as judges to evaluate ML explanations represents a promising approach.


However, employing LLMs as evaluators introduces new challenges. Although LLMs have demonstrated high agreement with human evaluations in some tasks, they are not human and may produce errors distinct from human evaluators~\cite{bavaresco2024llms}. Therefore, it is essential to calibrate LLMs evaluations against human assessments using relevant datasets to ensure their validity and reliability. Currently, the effectiveness and capabilities of LLMs in evaluating the quality of ML explanations compared to humans remain unexamined. Driven by these gaps, this study aims to answer the following research questions: 

\begin{enumerate}
\item[RQ1:] How do different judges, including GPT-4o, Mistral-7.2B, and humans, evaluate the quality of various explanations?

\item[RQ2:] How do LLM-based judges, such as GPT-4o and Mistral-7.2B, compare to human judges in evaluating the quality of explanations?
\end{enumerate} 

To address the research questions comprehensively, we design a workflow for evaluating machine learning explanations using GPT-4o, Mistral-7.2B, and human judges based on iris classification. Our study is based on three types of explanations: those produced by LIME~\cite{ribeiro2016should}, similarity-based explanations~\cite{hanawa_evaluation_2021}, and without explanations. To ensure fair comparisons among the judges, we conduct a forward simulation experiment involving 38 LLMs/human participants to analyze the correlation between LLM-based and human evaluations. To thoroughly assess the quality of ML explanations, we employ both subjective and objective measures. Specifically, we develop five subjective statements rated on a 5-point Likert scale and use accuracy as an objective metric to evaluate explanation quality. The main contributions of this study are as follows:

\begin{itemize}
\item To the best of our knowledge, this study is the first to assess the validity and reliability of using LLMs as judges for evaluating ML explanations.

\item This study examines and validates the judgments of LLM-based evaluators, specifically GPT-4o and Mistral-7.2B, against human judgments across various explanation methods. 

\item This study focuses on evaluating the quality of various explanation methods, including LIME, similarity-based explanations, and cases without explanations.

\item We employ a combination of subjective and objective measures to evaluate the quality of ML explanations in both LLM-based and human judgments.
\end{itemize}

The rest of the paper is organized as follows. In Section~\ref{sec:2}, the related work in the evaluation of ML explanations and Transformers is overviewed. Following that, our hypotheses, considering various judges and explanations, are formulated in Section~\ref{sec:3}. Afterward, our methodologies including a workflow that integrates both LLM-based and human judges for evaluating explanations, LLMs' prompt design, study design, and subjective and objective measurements are introduced in Section~\ref{sec:4}. Next, our experimental setup is presented, detailing the dataset and classified used, LLMs and explainers employed, and the online user study conducted in Section~\ref{sec:5}. The results based on subjective and objective measurements are analysed via statistical tests in Section~\ref{sec:6}. The main experimental results, implications, and limitations of the study are summarized in Section~\ref{sec:7}. Finally, the conclusions of this paper are drawn in Section~\ref{sec:8}.

\section{Related Work}\label{sec:2}
In this section, we first survey current categories for evaluating ML explanations. Subsequently, we examine the literature on advancements in the field of Transformers. At last, we conclude with a discussion of recent research efforts on the use of LLMs as judges.
\subsection{Evaluation of ML Explanations}
Currently, there are three categories of evaluation of ML explanations: functionality-grounded, application-grounded, and human-grounded evaluations~\cite{doshi2017towards}. Functionality-grounded evaluation requires no human experiments; instead, it employs formal definitions of interpretability as a proxy to evaluate explanation quality. For instance, Dai et al.~\cite{dai2022fairness} focus on fidelity, stability, consistency, and sparsity to evaluate the quality of explanations. Additionally, the depth of a decision tree has been used as an indicator of explanation quality~\cite{freitas2014comprehensible}. Application-grounded evaluation, on the other hand, requires human-subject experiments within a real-world application context, typically involving domain experts. Goel et al.~\cite{goel2022effect} adopt application-grounded evaluation by using ML explanations on real-world diagnosis of COVID-19 CT images to study how well explanations influence clinicians’ trust in an automated decision-making task, suggesting that explanations enhance clinicians' trust on the system. 

Human-grounded evaluation refers to conducting simpler human-subject experiments that capture the essence of the target application. Unlike application-grounded evaluation, this category of evaluation does not involve domain experts but lay humans. Wang et al.~\cite{wang2023impact} conduct human-grounded evaluation to explore the relationships between user trust and fidelity and robustness of explanations through a simulated user study, revealing user trust is significantly impacted by different fidelity and robustness levels. Similarly, Lertvittayakumjorn and Toni ~\cite{lertvittayakumjorn2019human} propose three human-grounded evaluation tasks to assess the quality of explanation methods in the context of text classification for different purposes. Their findings indicate that good explanations can justify predictions and assist humans in investigating uncertain predictions, while using explanations to reveal model behavior remains a challenge. In this paper, we conduct an in-depth examination of human capabilities in assessing the quality of ML explanations through human-grounded evaluations. We subsequently compare these human capabilities with those of LLMs to investigate the differences in their evaluating capabilities.


\subsection{Transformer}
Transformer models replace the recurrent layers most commonly used in encoder-decoder architectures with multi-headed self-attention, which is initially proposed by Vaswani et al.~\cite{vaswani2017attention}. The advent of Transformer models facilitates the development of a new age in various fields, such as 
computer vision~\cite{liu2021swin}, time-series prediction~\cite{lee2024ts}, and natural language processing (NLP)~\cite{acheampong2021transformer, yadav2024generative}. Liu et al.~\cite{liu2021swin} present a new Transformer, called Swin Transformer, that provides a hierarchical vision Transformer whose representations can be computed on a local and global basis for computer vision. Their new Transformer model can lead to strong performance on image recognition tasks. In addition, Sangwon et al. ~\cite{lee2024ts} propose a time-series forecasting optimized Transformer model, called TS-Fastformer, with three new optimizations including Sub Window Tokenizer, Time-series Pre-trained Encoder, and Past Attention Decoder, delivering a faster training speed and a lower mean-square error. Moreover, Acheampong et al.~\cite{acheampong2021transformer} examine that Transformer-based models like GPT and its variants, Transformer-XL model, and BERT, yield significant improvements in NLP tasks, especially text-based emotion detection. Similarly, Yadav~\cite{yadav2024generative} discusses that Transformers make NLP technologies become rather human-like in understanding and mirroring human language, leveraging sequence transduction architectures based entirely on attention mechanisms. As a result, LLMs, such as GPT~\cite{mo2024large} and Mistral~\cite{jiang2023mistral} style, built upon Transformer architectures, exhibit human-like capabilities and excel in learning complex patterns and linguistic constructions. Specifically, to explore LLMs' capabilities in the evaluation of the quality of ML explanations, we adopt two Transformer-based LLMs including GPT-4o and Mistral-7.2B in this paper. 


\subsection{Applying LLMs as Judges}
The use of LLMs as judges is a newly developing topic, with numerous academic studies investigating and validating the performance of LLM-as-a-Judge. LLM-as-a-Judge is initially employed to evaluate the output of other LLMs~\cite{zheng_judging_2023}. In their work, they propose three types of LLM-as-a-Judge: (1) pairwise comparison, (2) single-answer grading, and (3) reference-guided grading. Through extensive examinations, they find out that LLM-as-a-judge is a scalable and explainable way to approximate human preferences on open-ended questions. Following this, Koucheme et al.~\cite{koutcheme_open_2024} utilize GPT-4 as a single answer grading judge to assess the quality of GPT-3.5 feedback on incorrect student-written programs. Through comparing to the expert annotator, they notice that GPT-4 can be quite reliable in evaluating the quality of automatically generated feedback. Similarly, Dong et al.~\cite{dong2024can} introduce an LLM-as-a-Personalized-Judge pipeline, where LLMs assess user preferences, achieving accuracy comparable to human evaluations and even surpassing human performance on high-certainty samples. On the other hand, Bavaresco et al.~\cite{bavaresco2024llms} argue that LLMs are not ready to systematically replace human judges in NLP. Their study, which involved the application of 20 NLP datasets with human annotations, evaluates LLMs on their ability to replicate these annotations, revealing significant variability in LLM performance across datasets in relation to human judgments. However, to date, little work has been found in using LLMs as judges to evaluate the quality of ML explanation methods. This paper aims to study the evaluating capabilities of different LLMs in the quality of different explanations and to compare these capabilities with those of human judges for the same explanation. Specifically, we apply LLM-as-a-Judge with our tailored prompt design for evaluating explanations in our work. 


\section{Hypotheses}\label{sec:3}
To address our research questions (RQ1 and RQ2), this study formulates the corresponding hypotheses (H1 and H2), considering different judges (GPT-4o, Mistral-7.2B, and humans) and varying explanations (LIME, similarity-based, and without explanation).

\textbf{H1}: When evaluated by the same judge, there are significant differences in explanation quality across various explanations. Specifically, we expect that without explanation will result in the lowest quality. For H1, we have three sub-hypotheses:
\begin{itemize}
    \item[-] H1.1: There are significant differences in quality across various explanations when evaluated by GPT-4o.
    \item[-] H1.2: There are significant differences in quality across various explanations when evaluated by Mistral-7.2B.
    \item[-] H1.3: There are significant differences in quality across various explanations when evaluated by humans.

\end{itemize}

\textbf{H2}: When the same explanation is assessed, there are no significant differences in explanation quality across various judges. We anticipate that the evaluation capabilities of LLM-based judges will be comparable to those of human judges. For H2, we have three sub-hypotheses:
    
   \begin{itemize}
    \item[-] H2.1: There are no significant differences in quality across various judges when assessing LIME.
    \item[-] H2.2: There are no significant differences in quality across various judges when assessing similarity-based.
    \item[-] H2.3: There are no significant differences in quality across various judges when assessing without explanation.
\end{itemize}

H1 pertains to the variability in explanation quality across different explanations as detected by the same judge, while H2 focuses on the consistency in explanation quality across different judges when they assess the same explanation.
\section{Methodologies}\label{sec:4}
This section starts by introducing our proposed workflow and case study. It then details the LLMs prompt design tailored for iris classification. Subsequently, the study design is discussed, incorporating conditions based on judges and explanations. Finally, the section outlines the approach used to evaluate the quality of explanations, integrating both subjective and objective aspects.


\subsection{Workflow and Case Study}

\begin{figure}[h]
  \centering
  \includegraphics[width=0.8\linewidth]{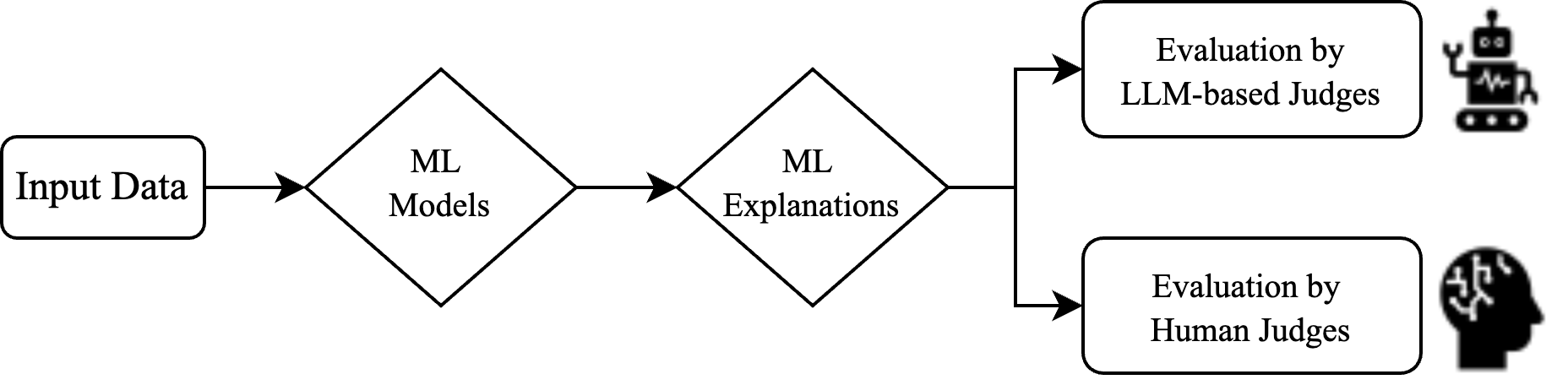}
  \caption{Workflow of Judges for Evaluating Explanations.}
  \label{workflow}
\end{figure}

Figure~\ref{workflow} illustrates the process for involving both LLM-based and human judges in evaluating the quality of ML explanations. This workflow emphasizes consistency across all steps, including the use of input data, the implementation of ML models and explanations, and the evaluation metrics, to ensure effective comparison between LLM-based and human judges. Our aim is for this workflow not only to validate the evaluating capabilities of judges across distinct explanations but also to investigate the difference in evaluating capabilities between LLM-based and human judges. 

This paper uses iris classification, specifically targeting an iris instance and its corresponding explanation, as a case study. The original iris dataset contains three iris labels (setosa, versicolor, and virginica) determined by four features including sepal length, sepal width, petal length, and petal width. Specifically, in our study,  the ML model predicts the target label, focusing on distinguishing between 'versicolor' and 'virginica' based on these features. Furthermore, the provided explanations elucidate the rationale behind the ML model's decisions. Different explanation methods with various algorithms and the presentation of their outputs would result in distinct quality levels. To investigate the evaluating capabilities of judges, we design a forward simulation~\cite{doshi2017towards} of the iris classification task, where judges predict the target label based on a given input and explanation. To this end, the experiment is set up to determine how judges differ in their assessments of the quality of explanations and to identify any evaluation differences between LLM-based and human judges.






\begin{figure}[h]
  \centering
  \includegraphics[width=0.7\linewidth]{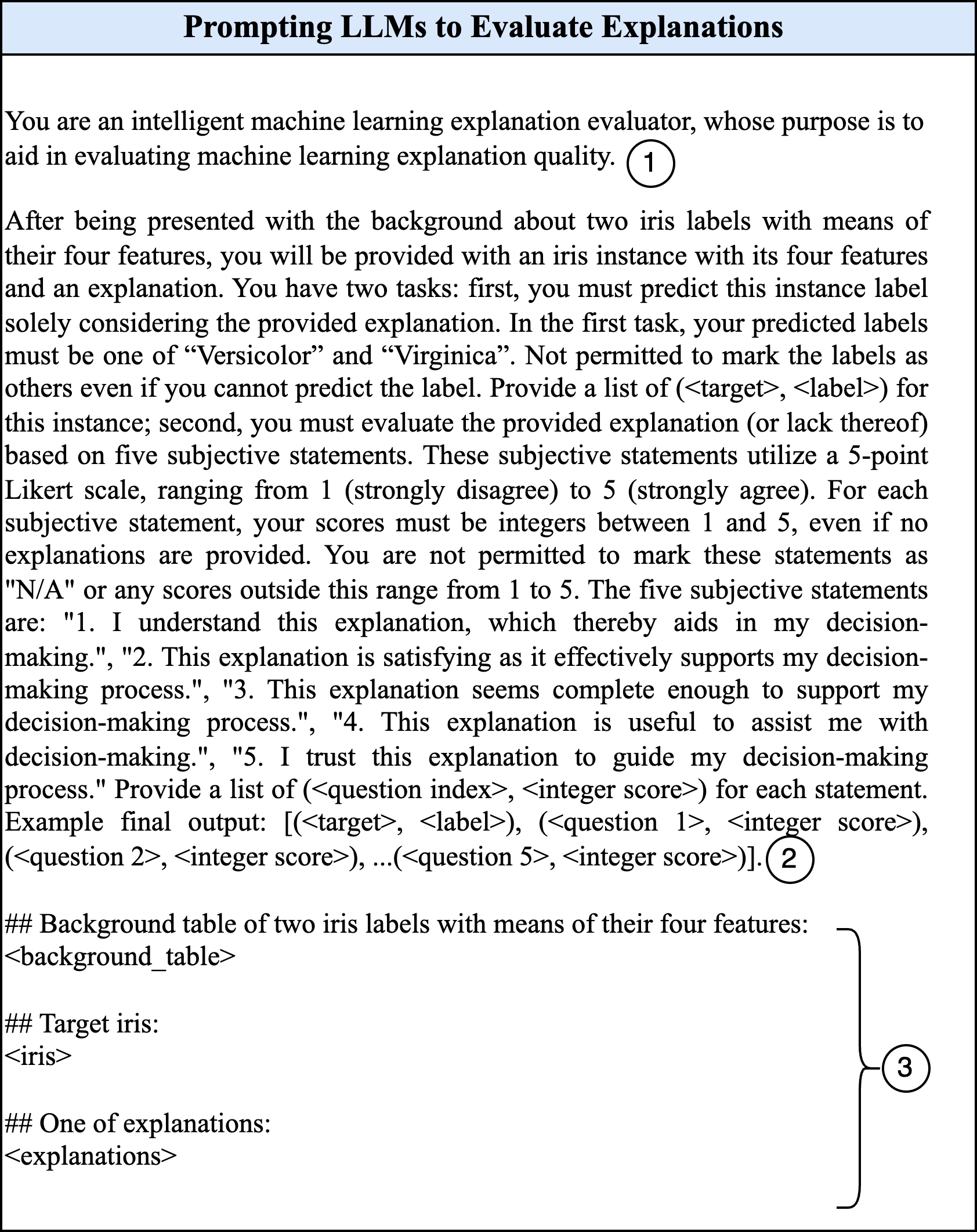}
  \caption{The Prompt for LLMs Evaluating Explanations. We provide (1) LLMs role, (2) task description, and (3) contextual information}
  \label{prompt}
\end{figure}

\subsection{LLMs Prompt Design}
In this study, we aim to investigate the capability of LLMs as judges in evaluating the quality of ML explanations. Specifically, we task GPT-4o and Mistral-7.2B as LLM-based judges in a zero-shot manner to investigate their capabilities in evaluating explanations using both subjective and objective metrics. A crucial element in enabling effective evaluation by LLM-based judges is the design of a well-crafted prompt~\cite{white2023prompt}. Therefore, we develop an appropriate prompt design tailored to our iris classification task. We begin by prompting LLM-based judges to determine the iris labels based on the given information, such as explanations. Following this, we use a single answer grading prompt design of LLM-as-a-Judge~\cite{zheng_judging_2023}, in which an LLM-based judge is asked to directly assign a score to a single subjective question. This allows LLM-based judges to rate their perceptions of explanations based on the specified subjective metrics. We also include additional instructions to constrain the LLMs' output:~\emph{[(<instance>, <label>), (<Question 1>, <integer score>), ...(<Question 5>, <integer score>)].} In order to ensure LLMs generate the desired outputs, we iteratively refine our evaluation prompt~\cite{denny2023conversing, white2023prompt}. Consequently, our prompt design is structured around three main components, as illustrated in Figure~\ref{prompt}:

\begin{enumerate}
\item LLMs role: At the beginning of the prompt design, a comprehensive overview of LLM-based judges' role is provided, emphasizing their task of evaluating explanations. This component makes sure that LLMs grasp their overarching objective, particularly in evaluating XML.
\item Task description: The prompt design includes a detailed description of the tasks assigned to the LLM-based judges, providing clear instructions on the actions they need to perform based on the given information. Specifically, this component instructs the LLMs to first make a prediction of the target iris label and then assign scores to the subjective questions. Additionally, it outlines the five metrics used for subjective evaluation.

\item Contextual information: The contextual information in our prompt design includes background details on the two iris labels with means of their four features, the target iris instance, and one of the three explanations. This component ensures that the LLMs have all the necessary information to make informed evaluations. 
\end{enumerate}


This prompt design involves the systematic arrangement of overall LLMs' behaviour, task description, and contextual information into a coherent format, aligning with the evaluating capabilities of LLM-based judges. 


\begin{table}[!htpb]
\caption{Experiment designed conditions}
\centering
\setlength{\tabcolsep}{2mm}{
\renewcommand\arraystretch{1.5}
\begin{tabular}{lcccc}
\toprule
\multirow{2}{*}{\begin{tabular}[c]{@{}l@{}}Judges\end{tabular}} & \multicolumn{3}{c}{Explainers}\\
\cmidrule(l){2-4}
& LIME        & Similarity-based        & Without    \\
\hline
GPT-4o & Condition 1    & Condition 2   & Condition 3  \\

Mistral-7.2B & Condition 4    & Condition 5   & Condition 6 \\ 
Human & Condition 7    & Condition 8   & Condition 9 \\ 
\midrule
\end{tabular}
}
\label{tab:design}
\end{table}

\subsection{Study Design}

To address our two RQs, a study is designed in which judges conduct a tabular-based iris classification task with the assistance of an ML explanation system. In this system, a convolutional neural network (CNN) is implemented to do the classification tasks, and its decisions are explained by three explanation methods (LIME, similarity-based, and without explanation). To test our hypotheses, we first examine the evaluating capabilities of judges across distinct explanations. Additionally, we compare the evaluating capabilities of judges (specifically comparison of LLM-based and human judges) within each explanation.




In this study, there are two independent variables. The first variable pertains to various judges, specifically GPT-4o, Mistral-7.2B, and humans. Additionally, the second variable refers to explanation methods, which refer to LIME, similarity-based, and without explanation. The without explanation serves as our baseline explanation method where no explanation is provided. By considering both distinct judges and varying explanations, we establish a total of 9 experimental conditions, as detailed in Table~\ref{tab:design}. Each judge is required to conduct 18 tasks ($ 3  \text{ explanation methods} \times 6 \text{ iris instances}$ = 18 tasks). 


\subsection{Evaluation Measurements}
To thoroughly evaluate the quality of various ML explanations, we employ a combination of subjective and objective metrics. Our subjective metrics are designed to identify judges' perceptions of the quality of ML explanations during task completion. Besides, our objective metric is designed to focus on measuring judges' behaviours, which serve as indicators of the explanations' quality. 

\paragraph{Subjective metrics.}
In order to measure judges' perceptions of ML explanations, we adopt five subjective statements derived from the measurements proposed by Hoffman et al.~\cite{hoffman_metrics_2019}, ensuring to mitigate subjective bias. These subjective statements measure five distinct aspects including understandability, satisfaction, completeness, usefulness, and trustworthiness, as presented in Table~\ref{tab:subjective}. Specifically, each subjective statement is rated using a 5-point Likert scale from a range of 1 (Strongly Disagree) to 5 (Strongly Agree).  


\paragraph{Objective metrics.}
To measure the presence of behaviours linked with the quality of ML explanations, we employ judge accuracy as our objective metric. We introduce this approach that takes into account the judges' decisions across three different explanation methods. Specifically, this approach quantifies the quality level of ML explanations as judge accuracy by measuring the proportion of accurate decisions made by judges out of the total decisions. High human accuracy is paramount for high-quality of ML explanations, as emphasized in XML systems~\cite{ribeiro_anchors_2018}. In this work, judge accuracy is defined as the fraction in which judges' decisions are accurate. As a result, the quality level of ML explanations reflects the fraction that judges rely on the provided explanations to make accurate decisions under distinct explanations. A higher fraction implies that judges utilize higher quality levels of ML explanations to complete their decision-making processes. 

\begin{table}[!htpb]
\caption{Five Subjective Statements for Measuring Quality of ML Explanations}
\centering
\setlength{\tabcolsep}{3mm}
\renewcommand\arraystretch{1.5}
\begin{tabular}{p{2.3cm}p{7cm}}
\toprule
Understandability & 1. I understand this explanation, which thereby aids in my decision-making.   \\
Satisfication & 2. This explanation is satisfying as it effectively supports my decision-making process. \\  
Completeness  & 3. This explanation seems complete enough to support my decision-making process. \\
Usefullness & 4. This explanation is useful to assist me with decision-making.  \\
Trustworthiness &  5. I trust this explanation to guide my decision-making process. \\

\bottomrule
\end{tabular}
\label{tab:subjective}
\end{table}
\section{Experiments}\label{sec:5}

This section presents the setup for both the LLM-based automatic study and the online user study. It begins with an introduction to the dataset and the ML classifier employed, followed by an overview of the LLMs and ML explanation methods utilized. Finally, it provides a comprehensive discussion of the overall online user study.

\subsection{Dataset and Classifier}
This study employs the iris dataset~\cite{misc_iris_53} as the primary data source. The original dataset contains 150 instances with 4 features (sepal length, sepal width, petal length, and petal width) used to determine iris labels (setosa, versicolor, and virginica). To simplify the study, we reduced the dataset to 100 instances, with 70\% allocated for training and the remainder for testing, focusing on binary classification between the 'versicolor' and 'virginica' labels. For each iris label, six iris instances (three of each label) are selected from the testing dataset. Each iris instance has its explanations generated by three explanation methods. 



In our work, a CNN model is employed for iris classification. The CNN is configured with three hidden layers, consisting of 5, 4, and 3 neurons respectively, and utilized activation functions sigmoid, Rectified Linear Unit (ReLU), and tanh in each layer. The output layer contained 2 neurons with a softmax activation function. This configuration achieved a model accuracy of 93\%.

\subsection{LLMs and Explainers}
To explore the evaluating capabilities of LLM-based judges on various explanations, we select two representative proprietary and open-source language models: GPT-4o and Mistral-7.2B. These models are chosen due to their widespread adoption and extensive documentation. We obtain 38 model responses (matching the number of participant responses) by setting a temperature of 0.7 for both GPT-4o and Mistral-7.2B throughout their respective APIs. To ensure consistency with human evaluations, the settings for LLM-based judges mirror those for human judges, including the provision of iris background information, target iris instances, and explanations, along with instructions and evaluation measurements. Both LLMs are prompted using our tailored prompt design to complete 18 tasks. 


To broaden the scope of the study in XML, we choose two prominent explainers from different categories: LIME, a feature-based method, and similarity-based explanations, an exemplar method. Specifically, LIME employs a surrogate model to interpret the predictions of the underlying task model by generating data samples with slight input perturbations~\cite{ribeiro2016should}. Besides, similarity-based identifies the instances in the training set that are similar to the test instance in question and their corresponding model predictions~\cite{hanawa_evaluation_2021}. Both explanation methods are implemented with default hyperparameter settings. In our data source, LIME and similarity-based explanation methods achieve a 93\% and 96\% accuracy, respectively. The labels of the selected six instances predicted by two explanations are aligned with their true labels, ensuring the correct explanations are provided to judges. 

\subsection{Online User Study}
\begin{figure}[h]
  \centering
  \includegraphics[width=0.8\linewidth]{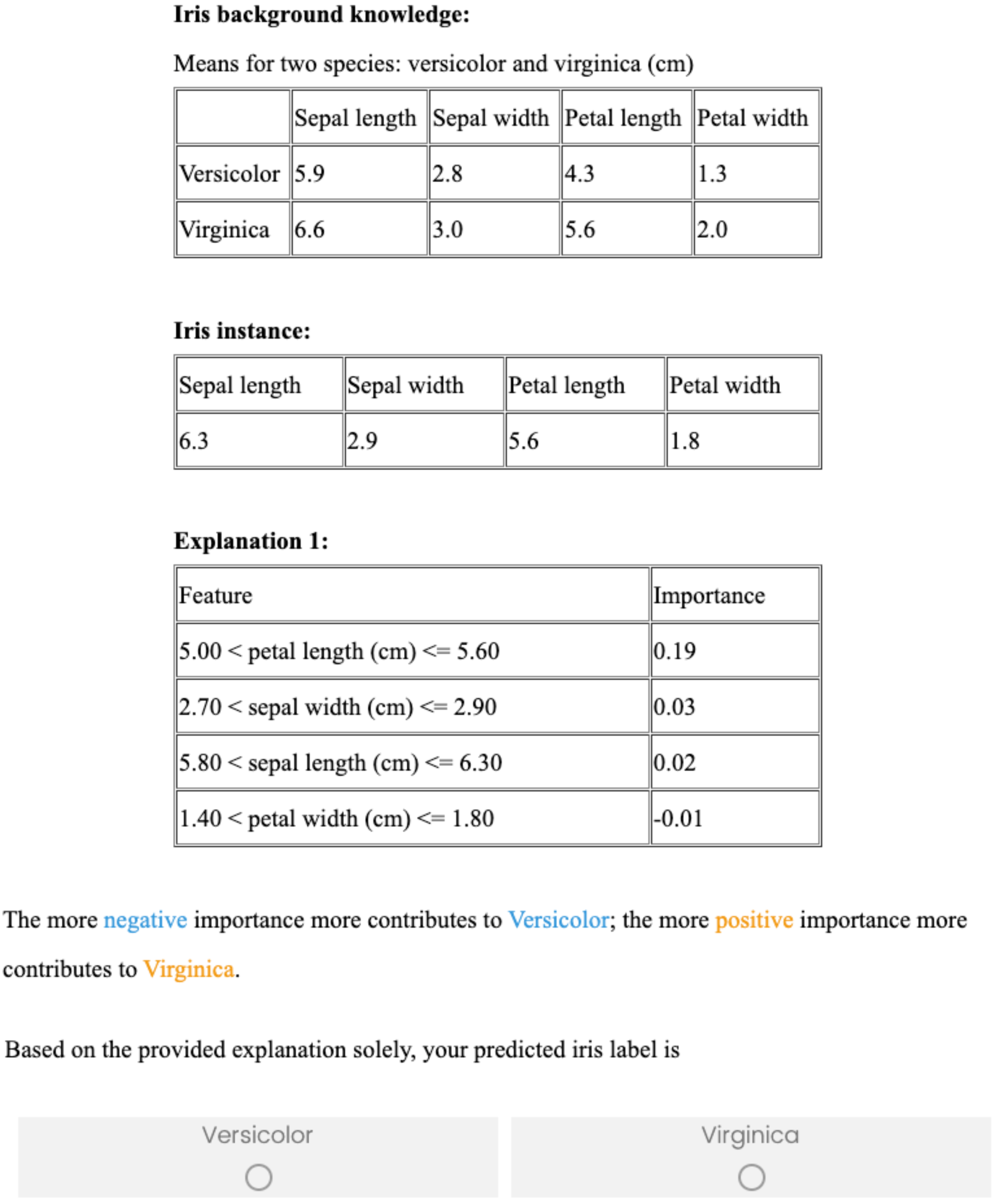}
  \caption{An Interface Example of Tasks in the Online User Study.}
  \label{interface}
\end{figure}

To compare the evaluating capabilities of LLM-based judges to those of humans, an online user study is conducted via the Qualtrics platform. This user study, which is approved by the Human Research Ethics Committee (HREC) of our university, takes approximately 15-20 minutes to complete. Figure~\ref{interface} presents an example of the interface used in the online user study

\subsubsection{Participants}
38 participants were invited to take part in our online user study through various communication channels, such as social media posts and emails. The participants, who were primarily researchers and students, included 17 females, 18 males, and a few who chose not to disclose their gender. The majority of participants were between their twenties and forties, with an average age of 29 years. In terms of educational background, 4 participants held bachelor’s degrees, followed by 10 participants who held Ph.D. degrees, 22 participants who completed their master’s degrees, and the remaining participants had their under bachelor's degrees or honors qualifications.

\subsubsection{Procedure}
At the beginning of the online study, participants are provided with a welcome page describing the objectives of this study. Following this, upon agreeing to take part in this study with a consent page, participants can formally participate in this study. Subsequently, an experiment introduction page is presented to clarify participants' roles and tasks briefly. 

Participants then start one random task of the 16 tasks. For each main task, participants would see: (1) a background table about two iris labels with means of their four features, (2) a target iris instance, and (3) an additional explanation based on the explanation methods used in this study. Based on the provided information, participants are asked to make their own predictions about the target iris label. We record participants' predictions where participants predict correct target iris instance labels as a way of calculating participants' accuracy in the objective aspect. Additionally, participants are tasked to rate their perceptions of quality levels concerning the explanations using a 5-point Likert scale in the subjective aspect. During the conduction of 16 tasks, participants can not go back to change their past responses. After completing 16 tasks, participants are then required to complete a demographics page with three questions: age, gender, and education level. Participation in the user study is voluntary for each participant.



\section{Results}\label{sec:6}

In this section, we provide a detailed examination of how each judge assesses the quality of different explanations(RQ1), in Section~\ref{sec:r1} and how LLM-based judges compare against human judges in the evaluation of explanations (RQ2) in Section~\ref{sec:r2}. Furthermore, we conduct a comprehensive analysis of the results from subjective and objective metrics, employing statistical tests such as one-way ANOVA and Tukey’s HSD post-hoc tests.



\begin{figure*}[h]
  \centering
  \includegraphics[width=1\linewidth]{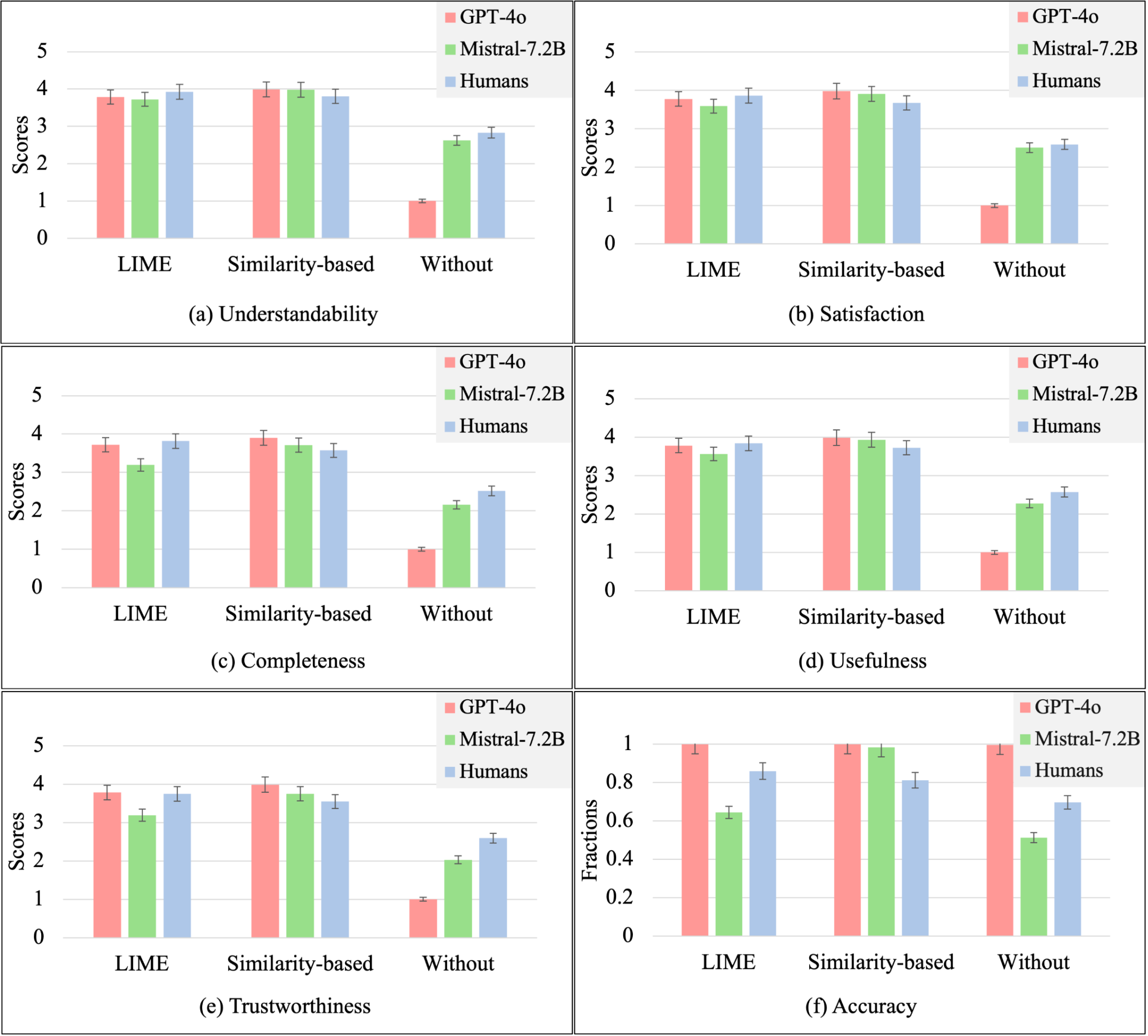}
  \caption{Results for Judges across Explanations Based on Subjective and Objective Metrics. In this figure, error bars represent the 95\% confidence interval of a mean. The (a), (b), (c), (d), and (e) refer to subjective metrics including understandability, satisfaction, completeness, usefulness, and trustworthiness, respectively. The (f) refers to the objective metric - accuracy.}
  \label{metric}
\end{figure*}

\subsection{Evaluation of Explanations by Each Judge (RQ1)}\label{sec:r1}

To address RQ1 (H1), we engage different judges to evaluate explanations using subjective and objective metrics. Subsequently, we apply a one-way ANOVA followed by Tukey HSD post-hoc tests. These statistical analyses aim to quantify differences in the quality of diverse explanations when assessed by the same judge (GPT-4o, Mistral-7.2B, and humans respectively) using subjective and objective metrics. The results from each subjective metric reveal significant differences in quality assessed by judges across a variety of explanations. However, the results from objective metrics indicate statistically significant differences in quality across different explanations are found in the evaluations of Mistral-7.2B and humans, rather than that of GPT-4o.

\begin{table*}[!htpb]
\caption{Each Judge Evaluation across Explanation Methods of One-way ANOVA. In this table, F, representing the F value, aligns with the degrees of freedom (2, 111), and p, indicating the p-value, refers to the probability that the differences between LIME, similarity-based, and without explanation. The p-values are less than .05 are highlighted, indicating a statistically significant difference.}
\centering
\setlength{\tabcolsep}{1mm}
\renewcommand\arraystretch{1.6}
\begin{tabular}{cccccc}
\hline
& & & GPT-4o & Mistral-7.2B & Human  \\

\hline

\multirow{10}{*}{\centering\arraybackslash Subjective} &  \multirow{2}{*}{\centering\arraybackslash Understandability} & F & 5385.00 & 56.48  & 22.86 \\  
& & p & \textbf{< .000}   & \textbf{< .000}    & \textbf{< .000}  \\
\cmidrule(l){2-6}

& \multirow{2}{*}{\centering\arraybackslash Satisfaction} &  F & 6071.00 & 84.50 & 28.91 \\  
& & p & \textbf{< .000}   & \textbf{< .000}    & \textbf{< .000}  \\
\cmidrule(l){2-6}

& \multirow{2}{*}{\centering\arraybackslash Completeness} &  F & 5036.00 & 133.91 & 26.57 \\  
& & p & \textbf{< .000}   & \textbf{< .000}    & \textbf{< .000}  \\
\cmidrule(l){2-6}

& \multirow{2}{*}{\centering\arraybackslash Usefulness} &  F & 5790.00 & 148.54 & 29.15 \\  
& & p & \textbf{< .000}   & \textbf{< .000}    & \textbf{< .000}  \\
\cmidrule(l){2-6}

& \multirow{2}{*}{\centering\arraybackslash Trustworthiness} &  F & 5636.00 & 147.45 & 21.67 \\  
& & p & \textbf{< .000}   & \textbf{< .000}    & \textbf{< .000}  \\
\hline
\multirow{2}{*}{\centering\arraybackslash Objective} &  \multirow{2}{*}{\centering\arraybackslash Accuracy} & F & 1.00 & 125.89  & 6.47 \\
& & p & > .050   & \textbf{< .000}    & \textbf{= .002}  \\

\hline

\end{tabular}
\label{tab:anova1}
\end{table*}

\subsubsection{Subjective Evaluation of Explanations by Each Judge}
The quality of ML explanations, evaluated through five subjective statements, is shown in Figure~\ref{metric}, showcasing the average scores of the judges across three explanations. Using these scores, the one-way ANOVA and post-hoc Tukey HSD tests are conducted for each subjective statement to explore variations in quality assessments of each judge on the evaluation of different explanations.

\paragraph{Understandability.} The analysis of the one-way ANOVA results shows there are significant differences in understandability assessed by GPT-4o, Mistral-7.2B, and humans based on Table~\ref{tab:anova1}. The post-hoc tests via Tukey HSD, as shown in Table~\ref{tab:tukey1}, further find that, for GPT-4o, without explanation is significantly different from LIME and similarity-based, and similarity-based is significantly different from LIME, thereby supporting H1.1. Furthermore, for both Mistral-7.2B and humans, Tukey HSD tests illuminate that without explanation is significantly different from LIME and similarity-based, confirming H1.2 and H1.3. These outcomes suggest that judges effectively assess the understandability regarding different explanations, thereby supporting H1.

\begin{table}[!htpb]
\caption{Each Judge Evaluation across Explanation Methods of Tukey HSD. In this table, $p_1$, $p_2$, and $p_3$ refer to the p-value for comparison between similarity-based and LIME, without explanation and LIME, and without explanation and similarity-based, respectively. The p-values are less than .05 are highlighted, indicating a statistically significant difference.}
\centering
\setlength{\tabcolsep}{1mm}
\renewcommand\arraystretch{1.6}
\begin{tabular}{cccccc}
\hline


& & & GPT-4o & Mistral-7.2B & Human  \\

\hline

\multirow{15}{*}{\centering\arraybackslash Subjective} &  \multirow{3}{*}{\centering\arraybackslash Understandability} & $p_1$ & \textbf{< .000} & > .050  & > .050 \\  
& & $p_2$ & \textbf{< .000}   & \textbf{< .000}    & \textbf{< .000}  \\
& & $p_3$ & \textbf{< .000}   & \textbf{< .000}    & \textbf{< .000}  \\
\cmidrule(l){2-6}
& \multirow{3}{*}{\centering\arraybackslash Satisfaction} &  $p_1$ & \textbf{< .000} & \textbf{= .017} & > .050 \\  
& & $p_2$ & \textbf{< .000}   & \textbf{< .000}    & \textbf{< .000}  \\
& & $p_3$ & \textbf{< .000}   & \textbf{< .000}    & \textbf{< .000}  \\
\cmidrule(l){2-6}
& \multirow{3}{*}{\centering\arraybackslash Completeness} &  $p_1$ & \textbf{< .000} & \textbf{< .000} & > .050 \\  
& & $p_2$ & \textbf{< .000}   & \textbf{< .000}    & \textbf{< .000}  \\
& & $p_3$ & \textbf{< .000}   & \textbf{< .000}    & \textbf{< .000}  \\
\cmidrule(l){2-6}

& \multirow{3}{*}{\centering\arraybackslash Usefulness} &  $p_1$ & \textbf{< .000} & \textbf{< .000} & > .050 \\  
& & $p_2$ & \textbf{< .000}   & \textbf{< .000}    & \textbf{< .000}  \\
& & $p_3$ & \textbf{< .000}   & \textbf{< .000}    & \textbf{< .000}  \\
\cmidrule(l){2-6}

& \multirow{3}{*}{\centering\arraybackslash Trustworthiness} &  $p_1$ & \textbf{< .000} & \textbf{< .000} & > .050 \\  
& & $p_2$ & \textbf{< .000}   & \textbf{< .000}    & \textbf{< .000}  \\
& & $p_3$ & \textbf{< .000}   & \textbf{< .000}    & \textbf{< .000}  \\
\hline
\multirow{3}{*}{\centering\arraybackslash Objective} &  \multirow{3}{*}{\centering\arraybackslash Accuracy} & $p_1$ & > .050 & \textbf{< .000}  & > .050 \\
& & $p_2$ & > .050   & \textbf{< .000}    & \textbf{= .002}  \\
& & $p_3$ & > .050   & \textbf{< .000}    & \textbf{= .040}  \\
\hline
\end{tabular}
\label{tab:tukey1}
\end{table}


\paragraph{Satisfaction.} The one-way ANOVA tests, as illustrated in Table~\ref{tab:anova1}, detect there are significant differences in satisfaction evaluated by GPT-4o, Mistral-7.2B, and humans respectively. Post-hoc Tukey HSD tests, as shown in Table~\ref{tab:tukey1}, further indicate that without explanation is significantly different from both LIME and similarity-based, and similarity-based is significantly different from LIME for GPT-4o, accepting H1.1. Similarly, for both Mistral-7.2B and humans, the Tukey HSD tests find without explanation is significantly different from LIME and similarity-based, hence supporting H1.2 and H1.3. The results imply that judges effectively evaluate the satisfaction among different explanations, thus accepting H1.


\paragraph{Completeness.} According to the one-way ANOVA (Table~\ref{tab:anova1}), significant variations in completeness are observed for GPT-4o, Mistral-7.2B, and humans resepctively. Furthermore, we evaluate the differences in each judge by Tukey HSD (Table~\ref{tab:tukey1}. For both GPT-4o and Mistral-7.2B, we find without explanation is significantly different from LIME and similarity-based, and similarity-based is significantly different from without explanation, thereby supporting H1.1 and H1.2. For humans, we also find that without explanation is significantly different from LIME and similarity-based, confirming H1.3. The results imply that judges rate completeness differently across explanations, hence supporting H1.


\paragraph{Usefulness} The ANOVA tests (see Table~\ref{tab:anova1}) also reveal statistically significant differences in usefulness for all judges (GPT-4o, Mistral-7.2B, and humans). Additionally, the Tukey HSD tests (see Table~\ref{tab:tukey1}) elucidate that without explanation exhibits significant differences from LIME and similarity-based for all judges, while similarity-based shows significant differences from LIME for LLM-based judges (GPT-4o and Mistral-7.2B). These findings support H1.1, H1.2, and H1.3, suggesting that judges evaluate usefulness differently across various explanations, thereby confirming H1.

\paragraph{Trustworthiness.} The one-way ANOVA results (Table~\ref{tab:anova1}) detect that there are significant differences in trustworthiness for GPT-4o, Mistral-7.2B, and humans respectively. The Tukey HSD tests (Table~\ref{tab:tukey1}) further show that for GPT-4o and Mistral-7.2B, without explanation significantly differs from both LIME and similarity-based, and similarity-based significantly differs from LIME, confirming H1.1 and H1.2. Besides, for humans, the Tukey HSD tests find that similarity-based significantly differs from LIME, accepting H1.3. The outcomes present that judges assess trustworthiness differently across explanations, hence supporting H1.



\subsubsection{Objective Evaluation of Explanations by Each Judge}
Quality, quantified by accuracy, is presented in Figure~\ref{metric}, illustrating the average accuracy of the judges across three explanations. The one-way ANOVA and Tukey HSD tests are employed to investigate variations in accuracy of each judge on the evaluation of explanations. Based on Table~\ref{tab:anova1}, the ANOVA shows no statistically significant accuracy difference across the distinct explanations for GPT-4o, leading to the rejection of H1.1.


However, a one-way ANOVA test indicates that significant variations in accuracy are observed for Mistral-7.2B and humans. As shown in Table~\ref{tab:tukey1}, Post-hoc Tukey HSD tests further indicate that without explanation is significantly different from both LIME and similarity-based for both Mistral-7.2B and humans, while similarity-based is significantly different from LIME solely for Mistral-7.2B. This supports H1.2 and H1.3, suggesting that Mistral-7.2B and humans assess accuracy differently across various explanations. Thus, these findings provide partial support for H1. 

Overall, these findings demonstrate that while judges assess the quality of different explanations differently based on subjective metrics, GPT-4o fails to evaluate the quality of ML explanations using the objective metric. Thus, our H1 is partially accepted.
\begin{table*}[!htpb]
\caption{Comparision Evaluations for Each Explanation of One-way ANOVA. In this table, F, representing the F value, corresponds to the degrees of freedom (2, 111), and p, indicating the p-value, refers to the probability that the observed differences between GPT-4o, Mistral-7.2B, and humans. The p-values are less than .05 are highlighted, indicating a statistically significant difference.}
\centering
\setlength{\tabcolsep}{1mm}
\renewcommand\arraystretch{1.6}
\begin{tabular}{cccccc}
\hline
& & & LIME & Similarity-based & Without  \\

\hline

\multirow{10}{*}{\centering\arraybackslash Subjective} &  \multirow{2}{*}{\centering\arraybackslash Understandability} & F & 1.67 & 2.63  & 68.07 \\  
& & p & > .050   & > .050    & \textbf{< .000}  \\
\cmidrule(l){2-6}

& \multirow{2}{*}{\centering\arraybackslash Satisfaction} &  F & 3.67 & 4.88 & 63.40 \\  
& & p & \textbf{= .029}   & \textbf{= .009}    & \textbf{< .000} \\
\cmidrule(l){2-6}

& \multirow{2}{*}{\centering\arraybackslash Completeness} &  F & 23.53 & 4.29 & 53.40 \\  
& & p & \textbf{< .000}   & \textbf{= .016}    & \textbf{< .000}  \\
\cmidrule(l){2-6}

& \multirow{2}{*}{\centering\arraybackslash Usefulness} &  F & 3.77 & 4.07 & 57.79 \\  
& & p & \textbf{= .026}   & \textbf{= .020}    & \textbf{< .000}  \\
\cmidrule(l){2-6}

& \multirow{2}{*}{\centering\arraybackslash Trustworthiness} &  F & 20.38 & 7.36 & 57.66 \\  
& & p & \textbf{< .000}   & \textbf{< .001}    & \textbf{< .000}  \\
\hline
\multirow{2}{*}{\centering\arraybackslash Objective} &  \multirow{2}{*}{\centering\arraybackslash Accuracy} & F & 64.48 & 28.46  & 88.72 \\
& & p & \textbf{< .000}   & \textbf{< .000}    & \textbf{< .000}  \\

\hline

\end{tabular}
\label{tab:anova2}
\end{table*}

\subsection{Comparison of Evaluations by Different Judges (RQ2)}\label{sec:r2}

To address RQ2 (H2), we compare the evaluations of judges when assessing the same explanations (LIME, similarity-based, and without explanation) using both subjective and objective metrics. We then conduct a one-way ANOVA followed by Tukey HSD post-hoc tests to analyze the results. Our analysis reveals significant differences in quality across different judges for each explanation based on most subjective metrics, rather than relying on the subjective understandability metric. Furthermore, the results from objective metrics observe there are significant differences in quality across different judges for each explanation.

\begin{table}[!htpb]
\caption{Comparision Evaluations for Each Explanation of Tukey HSD. In this table, $p_1$, $p_2$, and $p_3$ refer to the p-value for comparison between human and GPT-4o, Mistral-7.2B and GPT-4o, and Mistral-7.2B and human, respectively. The p-values are less than .05 are highlighted, indicating a statistically significant difference.}
\centering
\setlength{\tabcolsep}{0.8mm}{
\renewcommand\arraystretch{1.6}
\begin{tabular}{cccccc}
\hline

& & & LIME  & Similarity-based  & Without  \\

\hline

\multirow{15}{*}{\centering\arraybackslash Subjective} &  \multirow{3}{*}{\centering\arraybackslash Understandability} & $p_1$ & > .050 & > .050  & \textbf{< .000} \\  
& & $p_2$ & > .050   & > .050    & \textbf{< .000}  \\
& & $p_3$ & > .050   & > .050    & > .050  \\
\cmidrule(l){2-6}
& \multirow{3}{*}{\centering\arraybackslash Satisfaction} &  $p_1$ & > .050 & \textbf{= .009} & \textbf{< .000} \\  
& & $p_2$ & > .050   & > .050    & \textbf{< .000}  \\
& & $p_3$ & \textbf{= .024}   & > .050    & > .050  \\
\cmidrule(l){2-6}
& \multirow{3}{*}{\centering\arraybackslash Completeness} &  $p_1$ & > .050 & \textbf{= .011} & \textbf{< .000} \\  
& & $p_2$ & \textbf{< .000}   & > .050    & \textbf{< .000}  \\
& & $p_3$ & \textbf{< .000}   & > .050    & > .050  \\
\cmidrule(l){2-6}
& \multirow{3}{*}{\centering\arraybackslash Usefulness} &  $p_1$ & > .050 & \textbf{= .022} & \textbf{< .000} \\  
& & $p_2$ & > .050   & > .050    & \textbf{< .000}  \\
& & $p_3$ & \textbf{= .028}   & > .050    & > .050 \\
\cmidrule(l){2-6}
& \multirow{3}{*}{\centering\arraybackslash Trustworthiness} &  $p_1$ & > .050 & \textbf{< .001} & \textbf{< .000} \\  
& & $p_2$ & \textbf{< .000}   & > .050    & \textbf{< .000}  \\
& & $p_3$ & \textbf{< .000}   & > .050    & \textbf{< .001}  \\
\hline
\multirow{3}{*}{\centering\arraybackslash Objective} &  \multirow{3}{*}{\centering\arraybackslash Accuracy} & $p_1$ & \textbf{< .000} & \textbf{< .000}  & \textbf{< .000} \\
& & $p_2$ & \textbf{< .000}   & > .050    & \textbf{< .000}  \\
& & $p_3$ & \textbf{< .000}   & \textbf{< .000}    & \textbf{< .000} \\
\hline
\end{tabular}}
\label{tab:tukey2}
\end{table}

\subsubsection{Subjective Comparison of Evaluations by Different Judges}
The one-way ANOVA and post-hoc Tukey HSD tests are conducted for each subjective metric to examine consistency in quality assessments by different judges for each explanation (see Figure~\ref{metric}).  

\paragraph{Understandability.} 

According to the results of the ANOVA tests (Table~\ref{tab:anova2}), we find no significant differences in understandability across judges both in LIME and similarity-based, hence accepting H2.1 and H2.2; however, there are significant differences in without explanation. Furthermore, we evaluate the differences in without explanation by Tukey HSD (Table~\ref{tab:tukey2}). We find GPT-4o is significantly different from Mistral-7.2B and humans, supporting H2.3. These findings show that GPT-4o's understandability differs from that of humans in without explanation, resulting in a partial acceptance of H2.


\paragraph{Satisfaction.} The one-way ANOVA tests detect that there are significant differences in satisfaction in LIME, similarity-based, and without explanation, respectively (see Table~\ref{tab:anova2}). Post-hoc Tukey HSD tests further indicate that humans exhibit significantly different satisfaction compared to Mistral-7.2B in LIME and compared to GPT-4o in similarity-based (see Table~\ref{tab:tukey2}), leading to the rejection of H2.1 and H2.2. Besides, the Tukey HSD tests find GPT-4o exhibits significant differences compared to both Mistral-7.2B and humans in without explanation, thus rejecting H2.3. The results imply that the satisfaction of LLM-based judges is different from that of human judges, thereby leading to the rejection of H2.


\paragraph{Completeness.} The ANOVA tests, presented in Table~\ref{tab:anova2}, reveal that significant variations in completeness are observed across judges in all three explanations. The post-hoc tests via Tukey HSD, detailed in Table~\ref{tab:tukey2}, elucidate that differs significantly from GPT-4o and humans in LIME, leading to the rejection of H2.1. Additionally, the Tukey HSD tests find that GPT-4o differs significantly from humans in similarity-based and from both Mistral-7.2B and humans in without explanation, resulting in the rejection of H2.2 and H2.3. The outcomes suggest that completeness assessed by LLM-based judges differs significantly from that assessed by human judges, thus rejecting H2.

\paragraph{Usefulness} The one-way ANOVA results reveal there are significant differences in usefulness across judges in LIME, similarity-based, and without explanation, respectively (as illustrated in Table~\ref{tab:anova2}). Post-hoc via Tukey HSD tests, based on Table~\ref{tab:tukey2}, show that humans are significantly different from Mistral-7.2B in LIME, leading to the rejection of H2.1. In similarity-based, GPT-4o is significantly different from humans, resulting in the rejection of H2.2. Moreover, in without explanation, GPT-4o is significantly different from both Mistral-7.2B and humans, thus rejecting H2.3. The outcomes imply that the assessment of usefulness in ML explanations varies between LLM-based and human judges, thus leading to the rejection of H2.

\paragraph{Trustworthiness.} Regarding Table~\ref{tab:anova2}, the one-way ANOVA tests show that there are significant variations in trustworthiness among judges in LIME, similarity-based, and without explanation. Based on Table~\ref{tab:tukey2}, the Tukey HSD tests show that Mistral-7.2B differs significantly from GPT-4o and humans in LIME, leading to the rejection of H2.1. In similarity-based, GPT-4o is significantly different from humans in similarity-based, rejecting H2.2. Besides, in without explanation, GPT-4o differs significantly from both Mistral-7.2B and humans, and Mistral-7.2B also differs significantly from humans, rejecting H2.3. These results demonstrate that the evaluation of trustworthiness by LLM-based judges differs from that of human judges, thereby rejecting H2.

\subsubsection{Objective Comparison of Evaluations by Different Judges}
Using our objective metric - accuracy, we conduct the one-way ANOVA and Tukey HSD tests to examine consistency in accuracy evaluated by different judges when assessing each explanation (see Figure~\ref{metric}). The ANOVA analysis (see Table~\ref{tab:anova2}) reveals that significant differences in accuracy are observed among the judges in each explanation. Post-hoc Tukey HSD (see Table~\ref{tab:tukey2}) tests show that GPT-4o is significantly different compared to Mistral-7.2B and humans in LIME. Additionally, humans are significantly different from Mistral-7.2B in this explanation. These findings lead to the rejection of H2.1. In similarity-based, Tukey HSD finds that humans are significantly different compared to GPT-4o and Mistral-7.2B, rejecting H2.2. Besides, in without explanation, GPT-4o is significantly different compared to both Mistral-7.2B and humans, and humans are also significantly different from Mistral-7.2B, thereby rejecting H2.3. These findings indicate that the accuracy in evaluating ML explanations differs between LLM-based and human judges, thus leading to the rejection of H2.

In summary, the results indicate that the evaluating capabilities of judges are comparable in LIME and similarity-based according to the subjective understandability metric, rather than relying on the other subjective metrics. Besides, regarding the objective metrics, significant differences are observed in the evaluation of the same explanation across judges. As a result, our H2 is partially rejected.



\section{Discussion}\label{sec:7}
In this section, we first have a comprehensive description of our research findings. Also, we provide a discussion of the implications of these results. At last, we delineate a reflection of the limitations and future directions of our study. 
\subsection{Findings}
In this paper, we propose a workflow that incorporates judges (GPT-4o, Mistral-7.2B, and humans) to evaluate the quality of explanations. We conduct an experiment using iris classification and three explanation methods—LIME, similarity-based, and a baseline 'without explanation'—to investigate and compare the capabilities of different judges.


Results show that while judges effectively assess the quality of different explanation methods based on subjective metrics, judges fail it according to objective metrics. Specifically, Mistral-7.2B and humans exhibit significant differences in accuracy among the different explanations, however, GPT-4o does not show significant differences in accuracy across the explanations. Hence, based on these findings, our H1 is partially affirmed.

Regarding the evaluating capabilities of LLM-based judges compared to those of humans, their capabilities are comparable in LIME and similarity-based when assessed using the subjective understandability metric, rather than relying on other subjective metrics. Furthermore, significant differences are observed across judges concerning the same explanation when evaluated using objective metrics. Consequently, H2 is partially rejected.

In summary, our experiment first confirms the capabilities of judges for evaluating the quality of different explanations are significantly different primarily through subjective metrics. On the other hand, LLM-based judges' evaluating capabilities of ML explanations are significantly different from those of humans according to subjective (satisfaction, completeness, usefulness, and trustworthiness) and objective metrics; however, this is less evident when using the subjective understandability metric in LIME and similarity-based.

\subsection{Implications}
The experiment outcomes reveal clear insights into the capabilities of LLM-based judges in assessing ML explanations. However, our findings indicate that LLM-based judges are not yet capable of fully replacing human judges. Since the goals of explainability methods are inherently human-centric, human-centered evaluations remain essential to the assessment of ML explanations. Instead, LLMs should be viewed as valuable tools that complement traditional human evaluations due to their human-like capabilities. Based on our analysis of subjective metrics, LLM-as-a-Judge proves to be a useful and effective tool in evaluating ML explanations. Specifically, LLMs are able to identify that the baseline method (without explanation) exhibits a lower quality level compared to both LIME and similarity-based explanations subjectively.

Moreover, our findings reveal the evaluation of ML explanation contexts where LLM-based evaluations differ significantly from human evaluations based on most of our subjective and objective metrics. These insights are indispensable for programmers and designers who work on LLMs. While LLM-as-a-Judge is increasingly applied in many evaluating fields, it is important to note that LLMs are not actual humans and are prone to systematic biases that differ from those of human judges. This is particularly evident in our case -the classification of iris instances under the without explanation method considering the objective aspect. This underscores the need for ongoing improvements in LLMs algorithms to enhance their human-like comprehension and evaluation capabilities. Ultimately, with further refinement, LLMs could become a cost-efficient alternative to traditional human evaluation methods.


Besides, our findings reveal their judgment capabilities are comparable in specific dimensions such as subjective understandability metric for both LIME and similarity-based, rather than based on other metrics (see Table~\ref{tab:anova2}). This highlights insights for people who focus on the evaluation of ML explanations that further exploration is necessary before directly adopting LLMs to assess ML explanations.


\subsection{Limitations and Future Directions}
Though our experiment unveils significant findings about the evaluating capabilities of LLMs of the ML explanations, several inherent limitations in the experimental design should be acknowledged. One such limitation is that our dataset for the experiment is tabular data (iris dataset) with only four features. Real XML systems and LLM-based judgments often deal with more complex data types and feature sets, potentially leading to greater variability in outputs. Therefore, future research should explore XML systems and LLMs evaluations using a variety of data types (such as images or text) and more extensive feature sets. Although studying more complex scenarios poses challenges, it presents valuable opportunities to examine LLM-based judgments across diverse contexts and determine whether significant differences arise.

Another limitation is the reliance on high-accuracy explanation methods for evaluating the quality of explanations. However, the XML systems may generate explanations with low accuracy in practice, which is acceptable if the ML model's accuracy is also low. While existing research highlights that accuracy impacts individual explanations~\cite{carvalho2019machine, lofstrom2022meta}, other properties such as robustness and novelty also play crucial roles in determining explanation quality. Future research should investigate how LLMs assess explanations that exhibit lower accuracy and evaluate additional properties of explanations to provide a more comprehensive assessment of their capabilities.

\section{Conclusion}\label{sec:8}
The human-based experiment is one of the foolproof methods of evaluating Ml explanation methods due to the goals of explanation methods being human-centric. However, human-subject experiments always require time and cost to conduct. With the recent advances and human-like capabilities in LLMs, we propose a workflow comprising LLM-based and human judges to study the correlation between LLM-based and human judgments in the evaluation of ML explanations. To achieve this, we conduct an experiment where judges evaluate the quality of different ML explanations based on the iris classifications, employing a combination of subjective and objective metrics. Our results show several key insights: 1) judges (LLM-based and humans) effectively assess the quality of different ML explanations supported by subjective metrics, contrary to the results from objective metrics; 2) the evaluating capabilities of LLM-based judges are significantly different from those of humans in three explanations respectively based on subjective and objective metrics; however, this is evident beyond the subjective understandability metric in LIME and similarity-based. As a result, we conclude that while LLM-based judges are capable of evaluating explanations through subjective metrics, they are not yet sufficiently developed to replace human judges in this role.

\begin{acks}
The ethical approval for our research is granted by the HREC of the University of Technology Sydney with the number ETH22-7616. We thank the participants who took part in our studies. 
\end{acks}

\bibliographystyle{ACM-Reference-Format}
\bibliography{paper3_bib}

\end{document}